\documentclass[aps, a4paper, nofootinbib, showpacs, amsfonts]{revtex4}

\usepackage{amssymb,latexsym}
\usepackage{amsmath, amsthm}
\usepackage{amscd}
\usepackage{times}
\usepackage{epsfig}
\usepackage{psfrag}
\usepackage{graphicx}

\newcommand{\bea}{\begin{eqnarray}}
\newcommand{\eea}{\end{eqnarray}}
\newcommand{\be}{\begin{equation}}
\newcommand{\ee}{\end{equation}}

\begin{document}

\title{What a difference a quadrupole makes?}

\author{Daniele Malafarina}
\email{daniele.malafarina@nu.edu.kz}

\affiliation{Department of Physics, Nazarbayev University, 010000 Nur-Sultan, Kazakhstan}

\author{Sabina Sagynbayeva}
\email{sabina.sagynbayeva@stonybrook.edu}

\affiliation{Department of Physics, Nazarbayev University, 010000 Nur-Sultan, Kazakhstan}

\affiliation{Department of Physics and Astronomy, Stony Brook University, Stony Brook, NY 11794, USA}

\begin{abstract}
We consider some implications of the departure from spherical symmetry for static solutions of the vacuum Einstein's equations describing black hole mimickers. In particular, we investigate how the presence of mass quadrupole moment affects the collision energy of test particles. We show that collision processes in the vicinity of such exotic compact sources with non vanishing mass quadrupole moment may be significantly different from those in the vicinity of black holes. 
\end{abstract}



\maketitle

%%%%%%%%%%%%%%%%%%%%%%%%%%%%%%%

\section{Introduction}\label{intro}

The Kerr hypothesis states that every astrophysical black hole candidate in vacuum is described by the Kerr space-time and therefore must be completely characterized by two parameters only, namely mass and angular momentum.
The Kerr hypothesis is closely related to the No-Hair Theorem (NHT) which states that isolated, stationary, asymptotically flat black hole solutions in General Relativity (GR) are fully characterized by three quantities only, namely mass, angular momentum and charge
\cite{NH1, NH2}. These black hole solutions are: (i) Schwarzschild, describing static, uncharged black holes, (ii) Kerr, describing stationary, uncharged black holes, (iii) Reissner-Nordstrom, describing static, charged black holes and (iv) Kerr-Newman, describing stationary, charged black holes.  
For astrophysical sources the electric charge is usually considered to be negligible, which leaves Kerr as the only possibility for the geometry of rotating black hole candidates.
Of course, the above assumptions are not perfectly realized in nature 
and therefore the Kerr hypothesis remains an open problem in astrophysics.
In particular, real astrophysical objects are typically not isolated (being surrounded for example by matter fields and electromagnetic fields) and need not necessarily obey the symmetry of the Kerr space-time. 
This leads to the natural question of how to interpret the Kerr metric in the context of realistic astrophysical situations
(see for example \cite{gur, herrera}).
To what extent can we then rely on the Kerr metric to describe astrophysical black hole candidates? And what are the consequences of the departure from the black hole space-times for extreme astrophysical compact objects?

To investigate such questions it is worth considering the consequences of relaxing the symmetry requirements of the space-time
by considering the geometry of deformed sources. 
It is well known that there exist an infinite number of solutions of Einstein's equations describing stationary, axisymmetric and vacuum space-times which are not black holes and each solution can be obtained through a simple algorithmic procedure. In particular, restricting the attention to static and axially symmetric sources, the so called Weyl class \cite{Bonnor1992, weyl, weyl2}, there exists a one-to-one correspondence between solutions of Laplace equation and static, axially symmetric solutions of the vacuum field equations \cite{Quevedo1990, Pastora1994}. 
The question is then, what happens when a massive non spherical object collapses indefinitely under its own gravity? Since solutions belonging to Weyl's class, with the exception of Schwarzschild, posses curvature singularities at the location of the infinitely red-shifted surface (in Schwarzschild coordinates $r=2M$), we can envision two possible scenarios:
\begin{itemize}
    \item[(i)] All higher multipole moments are radiated away via gravitational waves during collapse and the space-time settles to the Schwarzschild geometry. In this case the horizon can be treated as a purely classical entity and the breakdown of the relativistic description occurs as the curvature diverges, i.e. in the vicinity of the central singularity at $r=0$.
    \item[(ii)] The collapsing object retains some of the higher multipole moments as the infalling matter approaches the surface $r=2M$. The horizon does not form and either matter is dispersed away or an exotic compact object forms. In this case, the curvature singularity appearing in non spherical solutions is naked, suggesting the possible existence of the boundary of a quantum-gravitational object. 
\end{itemize}

The first scenario leads to the formation of a black hole, while the second one leads to an exotic compact object. Both scenarios have advantages and disadvantages.
In case (i) we face the usual problems related to the formation of horizons during collapse and singularities as the endstate of collapse \cite{sing1,sing2}. The existence of horizons leads to problems such as, for example, the information loss paradox
\cite{inf}, while the resolution of such singularities may entail non local effects that alter the geometry in the vicinity of the horizon of the space-time, 
even if such horizon has low curvature and may be located far from the quantum-gravity dominated region (see \cite{malafarina} for a review).
In case (ii) we are also faced with the existence of singularities, 
in this case naked singularities that cause the breakdown of predictability at arbitrarily large distances.
However, the interpretation of such singularities is more straightforward, since they may just indicate that the boundary of the exotic compact object is located at a larger radius. In this case their eventual resolution may not lead to the problems mentioned above. 
In fact, in this case we may consider that the limits of applicability of the classical theory are reached already near the horizon, i.e. when $2GM/(c^2r)\simeq 1$. Massive particles do not cross the null divide and a quantum treatment, or a new theory of gravity, is necessary already at the horizon level.

In the following we will explore the consequences of the alternative point of view (ii). Hence, we will assume that not all higher multipole structure is radiated away during collapse and therefore the final exotic compact object forming from complete collapse will not be spherically symmetric. Consequently we will assume that a Weyl line element can describe the exterior geometry of such an extremely collapsed object, which need not be a black hole in a strict mathematical sense, and investigate how observations of the motion of test particles may allow to determine the nature of the geometry. 
It is worth noting that the boundary of such an object would be determined by the deviations from GR and the internal matter content of the object itself. In the case of quantum-gravity one would expect such a boundary to be very close to $r=2M$ (of the order of the Planck length) since the Planck length is the natural scale arising in quantum-gravity. However, since GR theory does not contain the limits of its applicability a new theory of gravity may in principle be needed at scales different from the Planck scale which would imply a larger boundary for the collapsed object. Therefore, all the considerations presented here for particle motion in the vicinity of $r=2M$ must be taken with a grain of salt, keeping in mind that the radius of the object's boundary is not known.
Despite being completely different from a black hole (i.e. having an actual surface and lacking the horizon) an exotic compact object that slightly departs from spherical symmetry may mimic the appearance of a black hole for certain kinds of observations. 
For example, the spectrum emitted from the accretion disk surrounding such a compact object may mimic that of a black hole (see \cite{orb1,orb2}).

In this perspective, we wish to investigate the effects that the absence of the horizon (and the presence of quadrupole moment) has on the appearance of the black hole mimicker. To this aim we will focus on two of the most important line-elements belonging to Weyl's class with non vanishing quadrupole moment, namely the Erez-Rosen (ER) metric \cite{ER1, armenti1977,ER2,ER3,ER4} and the Zipoy-Voothees (ZV) metric \cite{ZV1,ZV2,ZV3,ZV4}. 
Since in both cases the surface $r=2M$ in Schwarzschild-like coordinates has infinite redshift, the object may still appear like a black hole for certain observations, and may be labeled as a black hole mimicker. However, other kinds of observations may allow to distinguish the geometry from Schwarzschild and Kerr (see for example \cite{orb3,orb4,orb5,orb6}). 
In this work, we will concentrate the attention on the effect that the quadrupole moment has on the center of mass energy for the collision of test particles in the vicinity of the compact object. 
The study of collisions of test particles in the geometry of extreme compact objects was brought to prominence by Banados, Silk and West who considered the Schwarzschild and Kerr cases \cite{BSW}. Since then, many studies have been carried out to understand the mechanism of energy collisions near compact objects and its possible implications for astrophysics (see for example \cite{zav,coll1,coll2,coll3,coll4,coll5,coll6}).

The paper is organised as follows: In section \ref{axial} we will briefly review static axially symmetric vacuum solutions of Einstein's equations with particular attention to two important solutions with quadrupole moment, namely the ER and the ZV solutions. In section \ref{collisions} we will explore particle collisions in these space-times to investigate how they depart from the corresponding scenarios in the Schwarzschild geometry.
Finally in section \ref{discussion} we will discuss the implications that can be drawn for the physical validity of non black hole space-times as describing astrophysical compact objects. In the following we make use of geometrized units setting $G=c=1$.

\section{Static axially symmetric vacuum solutions}\label{axial}

The class of vacuum, static and axially symmetric space-times takes the name of Weyl class 
\cite{weyl}
and the most general line element in cylindrical coordinates $\{t,\rho,z,\phi\}$ takes the form
\be 
ds^2=-e^{2\lambda(\rho,z)}dt^2+e^{2\chi(\rho,z)-2\lambda(\rho,z)}(d\rho^2+dz^2)+\rho^2e^{-2\lambda(\rho,z)}d\phi^2 \; ,
\ee
where the functions $\lambda$ and $\chi$ need to be determined from Einstein's equations, that in this case become
\bea \label{laplace}
&&\lambda_{,\rho\rho}+\frac{\lambda_{,\rho}}{\rho}+\lambda_{,zz}=0 \; , \\ \label{quad}
&&\chi_{,\rho}=\rho(\lambda_{,\rho}^2-\lambda_{,z}^2) \; , \;\; \chi_{,z}=2\rho\lambda_{,\rho}\lambda_{,z} \; .
\eea 
It is easy to recognize that equation \eqref{laplace} is just Laplace equation in flat two dimensional space in cylindrical coordinates. Once a solution of equation \eqref{laplace} is given, equations \eqref{quad} are immediately integrated to give the second metric function thus solving the system completely. 
If we define
\bea
&&\alpha= \frac{R_++R_-}{2M} \; , \; \; \beta= \frac{R_+-R_-}{2M} \; , \\
&&R_\pm^2=\rho^2+(z\pm M)^2 \;,
\eea 
then solutions of equation \eqref{laplace} can be written as
\be \label{legendre}
\lambda_n=q_nP_n(\beta)Q_n(\alpha) \; , \; \; n=0,1,2...
\ee 
where $q_n$ are arbitrary constants, $P_n$ are the Legendre polynomials and $Q_n$ are Legendre functions of the second kind \cite{young}.
Therefore all solutions for static axially symmetric space-time are known and they are in one-to-one correspondence with solutions of Laplace equation \eqref{laplace} that can be given in terms of Legendre's polynomials (see \cite{Quevedo1990}). Furthermore, due to linearity of Laplace's equation, any linear combination of solutions $\lambda_n$ in \eqref{legendre} will also be a solution, with the non-linearity of Einstein's equations coming from equations \eqref{quad}. 

With the change of coordinates $\{\rho,z\}\rightarrow\{r,\theta\}$ given by
\bea 
\rho^2&=&\left(1-\frac{2M}{r}\right)r^2\sin^2\theta \; ,\\
z&=&(r-M)\cos\theta \; ,
\eea 
the most general line element of Weyl's class in Schwarzschild-like coordinates $\{t,r,\theta,\phi\}$ can be written as
\be \label{line}
ds^2=-F(r,\theta)dt^2+\frac{G(r,\theta)}{F(r,\theta)}\left[\frac{\Sigma(r,\theta)}{\Delta(r)}dr^2+\Sigma(r,\theta)r^2d\theta^2\right]+\frac{\Delta(r)}{F(r,\theta)}r^2\sin^2\theta d\phi^2 \; ,
\ee 
where 
\bea 
\Delta(r)&=&1-\frac{2M}{r} \; , \\ 
\Sigma(r,\theta)&=&1-\frac{2M}{r}+\frac{M^2}{r^2}\sin^2\theta\; ,
\eea 
and the metric functions which need to be determined from Einstein's equations now are $F(r,\theta)=e^{2\lambda}$ and $G(r,\theta)=e^{2\chi}$.
Therefore, in principle,  $F(r,\theta)$ and $G(r,\theta)$ can be determined once a solution of Laplace equation is given and therefore all exact solutions in the form \eqref{line} are known. However, while the physical interpretation of solutions of Laplace equations is straightforward in Newtonian mechanics not the same can be said for their relativistic counterparts. In fact, one may write the solution to Laplace equation in terms of a sum of Legendre polynomials and the coefficients $a_n$ multiplying each polynomial, which are usually called Weyl multipoles, do not correspond to the gravitational mass multipoles $M_n$
\cite{Pastora1994}. For example, the Weyl monopole is the Curzon solution \cite{curzon}, while the gravitational monopole is of course the Schwarzschild solution. Therefore the physical interpretation of Weyl's solutions is not straightforward.

Here we are interested in the role played by small departures from spherical symmetry in the form of non vanishing quadrupole moment for static solutions. 
The study of the multipolar structure of asymptotically flat solutions of Einstein's equations was done by Geroch \cite{multipoles1}, Hansen \cite{multipoles2}, Thorne \cite{multipoles3} and others (see for example \cite{multipoles4}). In brief we can say that there exist two sets of distinct multipole moments related to mass and angular momentum. For static solutions the angular momentum multipoles all identically vanish. Therefore in this work we are concerned only with the mass multipoles, and  the only vacuum static black hole (i.e. Schwarzschild) is defined by its mass monopole while all higher multipole moments vanish.
In this respect the two most interesting axially symmetric static solutions are the Erez-Rosen (ER) solution \cite{ER1}, which extends the Schwarzschild solution with the introduction of a full spectrum of non vanishing mass multipole moments depending on one parameter $q$ while keeping the mass monopole equal to the parameter $M$, and the Zipoy-Voorhees (ZV) solution \cite{ZV1,ZV2}, also known as $\gamma$-metric, which extends the Schwarzschild solution with the introduction of non vanishing even multipole moments of every order, all depending on a single parameter $\gamma$. 

Both solutions present a naked singularity at the surface $r=2M$, where in the spherical case the horizon is located. 
This is true in general for line elements belonging to Weyl's class but the nature and strength of the singularity may vary significantly depending on the solution and the values of the parameters involved (see for example \cite{semerak}). For example, both the prolate ER and the ZV metrics exhibit repulsive effects in the proximity of the singularity for certain values of the quadrupole moment showing that in such cases the singularity is 'weak' (see \cite{bini2,orb5}). Furthermore, for the ZV metric it was shown that the singularity is string-like in the prolate case and becomes point-like in the oblate case \cite{kodama}.

As said, we can understand the singularities as indicating the failure of the classical relativistic theory and hence the solutions would describe the exterior field of an exotic compact object with boundary $r_b>2M$, where the exact value of the boundary would depend on the new theory of gravity used to describe the source and thus can not be determined within GR.
It should be noted that in order for such solutions to be physically viable as the exterior field of an exotic compact object with boundary slightly larger than $r=2M$ such object should be stable against various types of perturbations (such as gravitational, electromagnetic, etc.) \cite{Teukolvsky}. However, due to the lack of a theory describing the matter source of such compact objects and the difficulties arising from axial symmetry, we are not able to investigate the stability here, and the issue is left open for future studies.

\subsection{Schwarzschild}
The Schwarzschild solution is spherically symmetric and is the only asymptotically flat, static and vacuum solution which describes a black hole.
Its metric function $\lambda$ is written in the form of equation \eqref{legendre} for $n=0$ as
\be 
\lambda_0=-P_0(\beta)Q_0(\alpha)=\frac{1}{2}\ln\frac{\alpha-1}{\alpha+1}\; .
\ee 
Accordingly, it can obtained from the line element \eqref{line} when $F$ and $G$ are given by
\bea 
F(r)&=&\Delta(r) \; , \\
G(r,\theta)&=&\frac{\Delta(r)}{\Sigma(r,\theta)} \; .
\eea

For Schwarzschild's mass multipole moments we have
\bea 
M_0&=&M\; , \\
M_{n}&=&0\; \; \text{ for } \; \; n\geq 1 .
\eea 
As it is well known the surface $r=2M$ identifies the event horizon and, in Schwarzschild's coordinates, it is a coordinate singularity. The curvature is finite at the horizon and the coordinate singularity can be removed by a suitable change of coordinates. Then trajectories may extend inside the horizon and every particle geodesic entering the same will terminate at the curvature singularity located at $r=0$. 

\subsection{Erez-Rosen}

The ER solution is the natural extension of the Schwarzschild metric with the addition of quadrupole moment and it is constructed from the sum of the first two non trivial solutions of Laplace's equation, i.e. $\lambda_n$ given in equation \eqref{legendre} with $n=0$ and $n=2$\footnote{Notice that, as expected, for GR the dipole contribution $n=1$ vanishes by making a suitable choice for the origin of the coordinates.}. Then we have
\be 
\lambda_{ER}=\lambda_0+q\lambda_2=-P_0(\beta)Q_0(\alpha)-qP_2(\beta)Q_2(\alpha)\; .
\ee 
In Schwarzschild-like coordinates the metric function $\lambda$ becomes
\be
F(r,\theta)= \Delta(r) e^{2q\Psi(r,\theta)}\; ,
\ee 
with
\be \label{psi}
\Psi(r,\theta)= \frac{3\cos^2\theta-1}{4}\left[\frac{3r^2-6Mr+2M^2}{2M^2}\ln\left(1-\frac{2M}{r}\right)+3\frac{r-M}{M}\right] \; .
\ee
The complete expression for $G$ is \cite{ER3}
\be
G(r,\theta)=\frac{\Delta(r)}{\Sigma(r,\theta)}e^{2q\Phi(r,\theta)} \; ,
\ee
with
\bea \nonumber
\Phi(r,\theta)&=&-3+\frac{2+q}{2}\ln\left(\frac{\Delta(r)}{\Sigma(r,\theta)}\right)-\frac{3(r-M)}{2M}\ln\Delta(r)+\frac{9q}{16}\sin^2\theta\left\{\left(\frac{r-M}{M}\right)^2(1-9\cos^2\theta)+\right. \\ \nonumber
&&+4\cos^2\theta-\frac{4}{3}+\left(\frac{r-M}{M}\right)\left[\left(\frac{r-M}{M}\right)^2(1-9\cos^2\theta)+7\cos^2\theta-\frac{5}{3}\right]\ln\Delta(r)+ \\  
&&+\left.\left(\frac{r^2-2Mr}{4M^2}\right)\left[\left(\frac{r-M}{M}\right)^2(1-9\cos^2\theta)-\sin^2\theta\right]\ln^2\Delta(r)\right\} \; .
\eea 
The monopole and quadrupole moments of the ER solution are
\bea 
M_0&=&M\; , \\
M_2&=& \frac{2qM^3}{15} \; .
\eea
Notice that for the ER solution we have that the mass parameter is given by the monopole moment $M_0$ and for $q=0$ we retrieve the Schwarzschild line element. Then cases where $q>0$ (and $q<0$ respectively) are interpreted as representing the field outside prolate (oblate) sources.
The ER metric has a true curvature singularity at $r=2M$ for $q\neq 0$.

\subsection{Zipoy-Voorhees}
The ZV solution, also known as $\gamma$-metric, is a `natural' axially symmetric extension of the Schwarzchild solution as it depends on only one extra parameter $\gamma$ describing the departure from spherical symmetry. 
By noticing that Schwarzschild is obtained from equation \eqref{legendre} by taking $q_0=-1$, it is straightforward to extend the solution to the ZV metric with
\be 
\lambda_{ZV}=-\gamma P_0(\beta)Q_0(\alpha)=\frac{\gamma}{2}\ln\frac{\alpha-1}{\alpha+1}\; .
\ee 
The metric functions in Schwarzschild-like coordinates are given by
\bea \label{zv-gtt}
F(r)&=&\Delta(r)^\gamma \; , \\ \label{ZV-grr}
G(r,\theta)&=&\left(\frac{\Delta(r)}{\Sigma(r,\theta)}\right)^{\gamma^2} \; ,
\eea 
and its monopole and quadrupole moments are
\bea \label{ZVM0}
M_0&=&M\gamma\;, \\ \label{ZVM2}
M_2&=&(1-\gamma^2)\frac{\gamma M^3}{3} \; .
\eea
Notice that for the ZV solution we have that all even multiple moments $M_{2n}$ do not vanish if $\gamma\neq 1$.
For $\gamma=1$ the line element reduces to Schwarzschild and cases where $\gamma>1$ (and $\gamma<1$ respectively) are interpreted as representing the field outside oblate (prolate) sources. Similarly to the ER metric, the ZV metric also has a true curvature singularity at $r=2M$ for $\gamma\neq 1$.

Note that other solutions characterized by the presence of quadrupole moment do exist. The structure of multipole moments of Weyl's solutions was extensively studied in \cite{Pastora1994} with the aim of characterizing solutions based on the properties of the source. Other notable solutions with quadrupole moment were studied for example in \cite{HP} and \cite{gutsunayev} and the role played by quadrupole moment in the motion of test particles was first considered in \cite{armenti1972}.

\section{Particle collisions}\label{collisions}

To describe the process of energetic collisions of two test particles of masses $m_1$ and $m_2$ in an axially symmetric space-time we follow the framework outlined by Banados, Silk and West in \cite{BSW}. 
A general framework for particle collisions was introduced in \cite{zav}, where a generic stationary line element with five metric functions of $r$ and $\theta$ was considered. The case of Weyl's metrics can be considered an important subclass of the above mentioned case.
The 4-velocity of each particle is given by $u^\mu_i=\{\dot{t}_i, \dot{r}_i, \dot{\theta}_i, \dot{\phi}_i\}$, where $i=1,2$ and the dot represents derivatives with respect to the proper time $\tau$ parametrizing the particle's trajectory. In the following we will restrict the attention to motion in the equatorial plane, thus setting $\theta=\pi/2$ and $\dot{\theta}=0$.

\begin{figure}[hhh]
\centering
\begin{minipage}{.45\textwidth}
\centering
\includegraphics[width=\columnwidth]{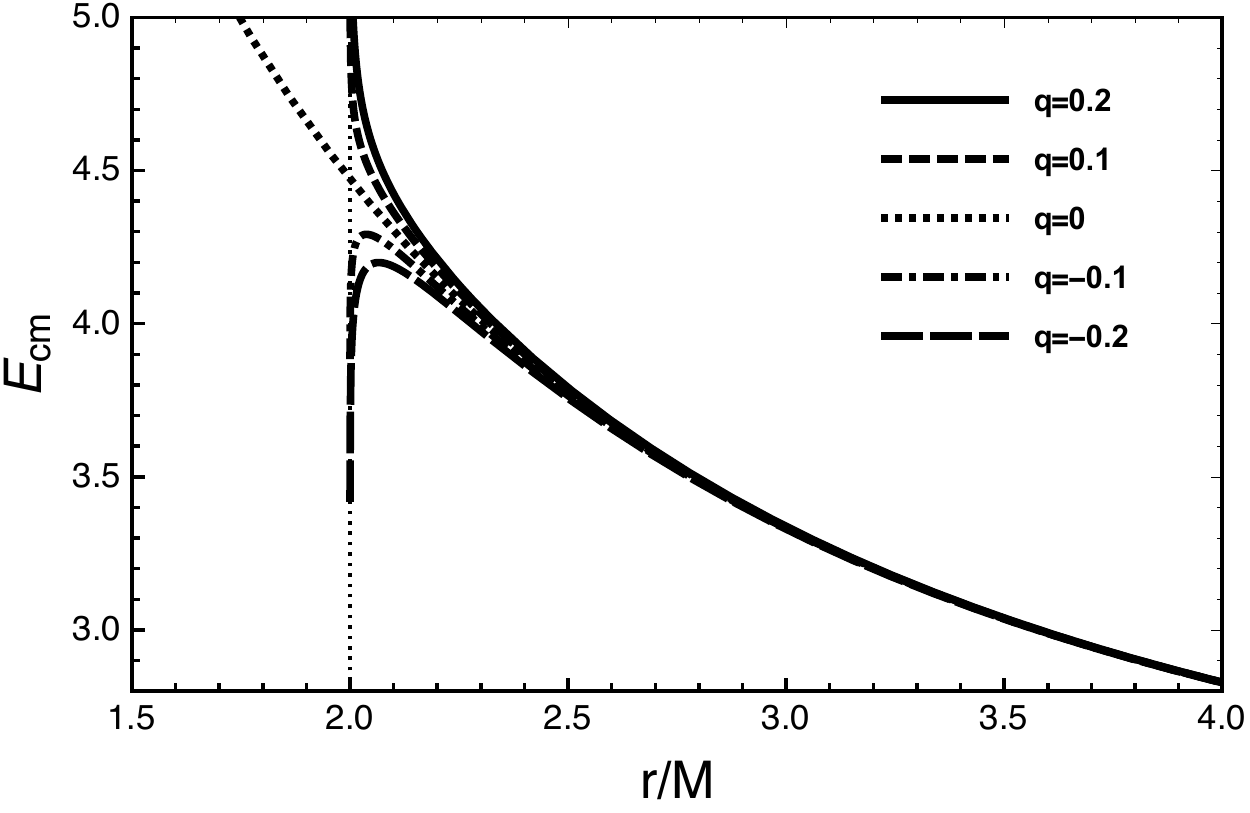}
\end{minipage}
%\hfill
\begin{minipage}{.45\textwidth}
\centering
\includegraphics[width=\columnwidth]{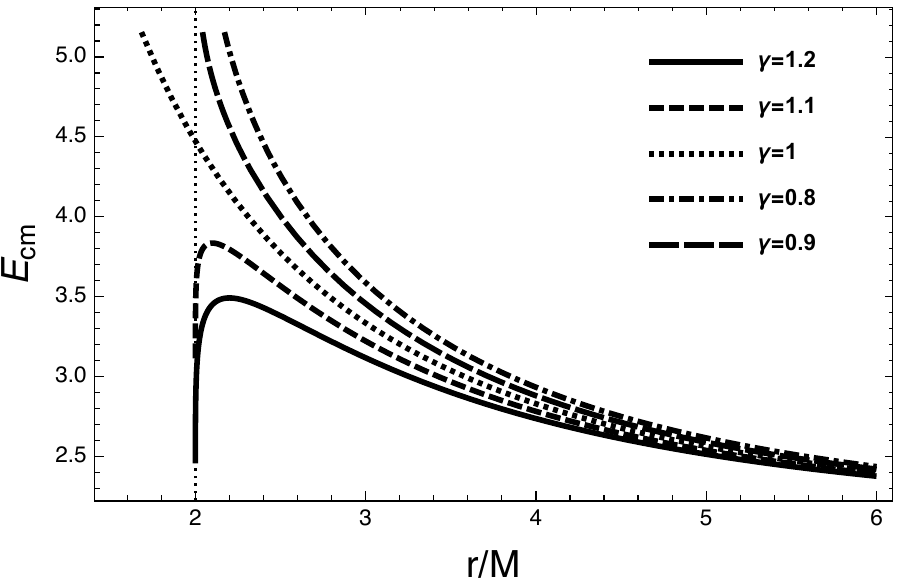}
\end{minipage}
\caption{The center of mass energy for the collision of two particles with given angular momentum $L_1=-L_2=4$ in the ER space-time (left panel) and in the ZV space-time (right panel). The Schwarzschild case is retrieved for $q=0$ in the ER metric and $\gamma=1$ in the ZV metric.}
\label{fig1}
\end{figure}

Given the symmetry of the space-time we know that there exist two Killing vectors, one related to time translations and one related to rotations about the symmetry axis, which correspond to two conserved quantities, namely the energy of the test particle $E$ and the component of its angular momentum along the direction of the symmetry axis $L$. Therefore, for each particle, omitting the subscript identifying the particle, from the line element \eqref{line} we have
\bea 
\dot{t}&=&\frac{E}{F}\; , \\
\dot{\phi}&=&\frac{F}{\Delta}\frac{L}{r^2} \; ,
\eea 
and the 4-velocity of the test particle in the equatorial plane becomes $u^\mu=\{E/F, \dot{r}, 0, (FL)/(\Delta r^2)\}$. Then  we can write
\be \label{p}
p_\mu p^\mu=-m^2=-\frac{E^2}{F}+\frac{G\Sigma}{F\Delta}\dot{r}^2+\frac{F}{\Delta}\frac{L^2}{r^2}\; .
\ee 
which can be used to obtain $\dot{r}^2$.
The center of mass energy $E_{\rm cm}$ for two colliding particles is given in general by
\be 
E_{\rm cm}^2=m_1^2+m_2^2-2g_{\mu\nu}p_1^\mu p_2^\nu \; ,
\ee 
where $p^\mu_i=m_iu^\mu_i$ (with $i=1,2$) is the particle's 4-momentum.
In the following, for the sake of clarity, we will consider particles with the same mass $m_1=m_2=m$ that have zero radial velocity at spatial infinity. 
Taking $\dot{r}\rightarrow 0$ for $r\rightarrow +\infty$, given the asymptotic flatness of the space-time, implies that the energy of the particle is $E=m$. Further we will consider the energy and angular momentum per unit mass, thus replacing $E$ and $L$ with $E/m$ and $L/m$ so that the expression for the center of mass energy per unit mass becomes
\be \label{Ecm}
\frac{E_{\rm cm}^2}{2}=1+\frac{1}{F}-\frac{F}{\Delta}\frac{L_1L_2}{r^2}-\frac{1}{F}\sqrt{1-F-\frac{F^2}{\Delta}\frac{L_1^2}{r^2}}\sqrt{1-F-\frac{F^2}{\Delta}\frac{L_2^2}{r^2}}  \; .
\ee 
This expression depends on $\Delta$ and involves only on one metric function, $F$ defining which solution we are describing. In the case of Schwarzschild (i.e. for $F=\Delta$) we retrieve the result obtained in \cite{BSW} for which the center of mass energy at the horizon is finite and is given by
\be 
\lim_{r\to 2M} E_{\rm cm}^{q=0}=\lim_{r\to 2M} E_{\rm cm}^{\gamma=1}=\frac{\sqrt{L_1^2-2L_1L_2+L_2^2+16M^2}}{2M}\; .
\ee 
On the other hand, for both the ER and ZV metrics we obtain that $E_{\rm cm}$ diverges for prolate sources (i.e. $q>0$ for ER and $\gamma<1$ for ZV, respectively)
\be 
\lim_{r\to 2M} E_{\rm cm}^{q>0}=\lim_{r\to 2M} E_{\rm cm}^{\gamma<1}=+\infty \; .
\ee
For oblate sources (i.e. $q<0$ for ER and $\gamma>1$ for ZV), the center of mass energy tends to a minimum value that is independent of the value of the quadrupole moment
\be 
\lim_{r\to 2M} E_{\rm cm}^{q<0}=\lim_{r\to 2M} E_{\rm cm}^{\gamma>1}=2 \; .
\ee
However, it must be noted that $E_{\rm cm}(r)$ for oblate sources approaches the minimum value at $r=2M$ with vertical tangent (i.e. diverging first derivative) and thus also in this case the center of mass energy is not defined for $r<2M$.
The behaviour of $E_{\rm cm}$ as a function of $r$ in the ER metric for different values of $q$ is illustrated in the left panel of figure \ref{fig1}, while the corresponding behaviour in the ZV metric for different values of $\gamma$ is illustrated in the right panel of figure \ref{fig1}. 
This shows that, under the assumption that the geometry describes a compact object in a theory of gravity beyond GR, with boundary slightly larger that $r=2M$, the center of mass energy for particle collisions in the proximity of the object's surface may be able to distinguish whether the geometry is that of a black hole or it has a non vanishing quadrupole moment.
	
\begin{figure}
	\includegraphics[width=0.48\columnwidth]{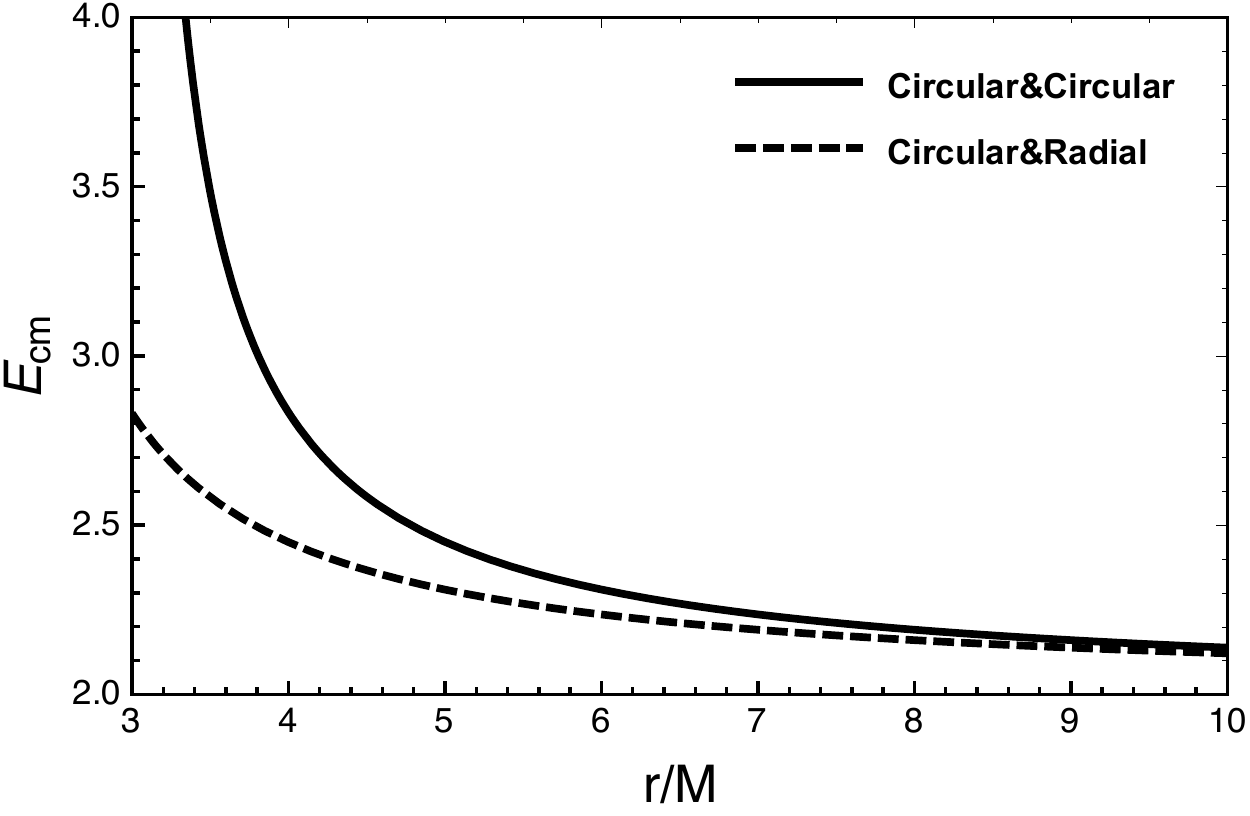}
	\includegraphics[width=0.48\columnwidth]{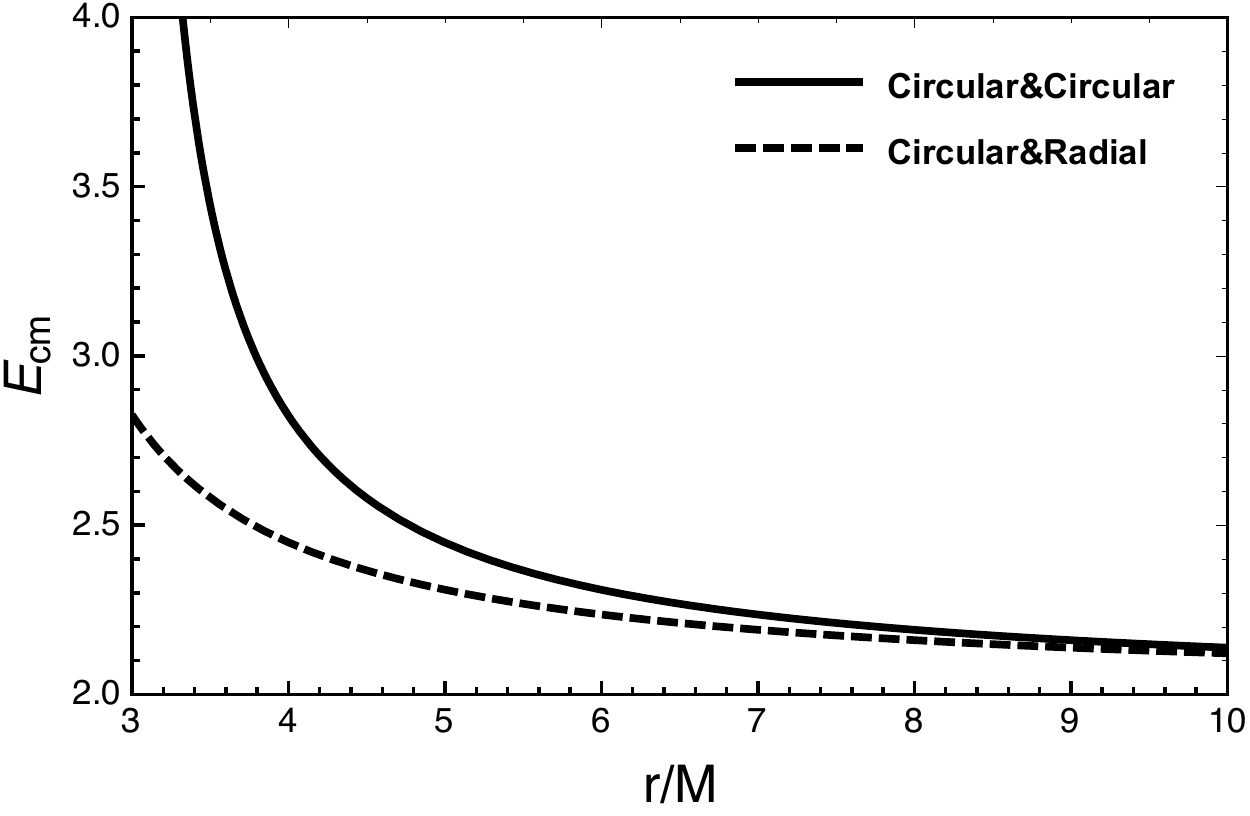}
    \caption{Collision of particles in the ER space-time. Solid lines represent both particles on circular orbits; dashed lines represent one particle is on circular orbits and one particle falling radially from infinity. Left panel: Oblate case ($q=-0.1$). Right panel: Prolate case ($q=0.1$).}
    \label{fig:ERcirc11}
\end{figure}

\begin{figure}
	\includegraphics[width=0.48\columnwidth]{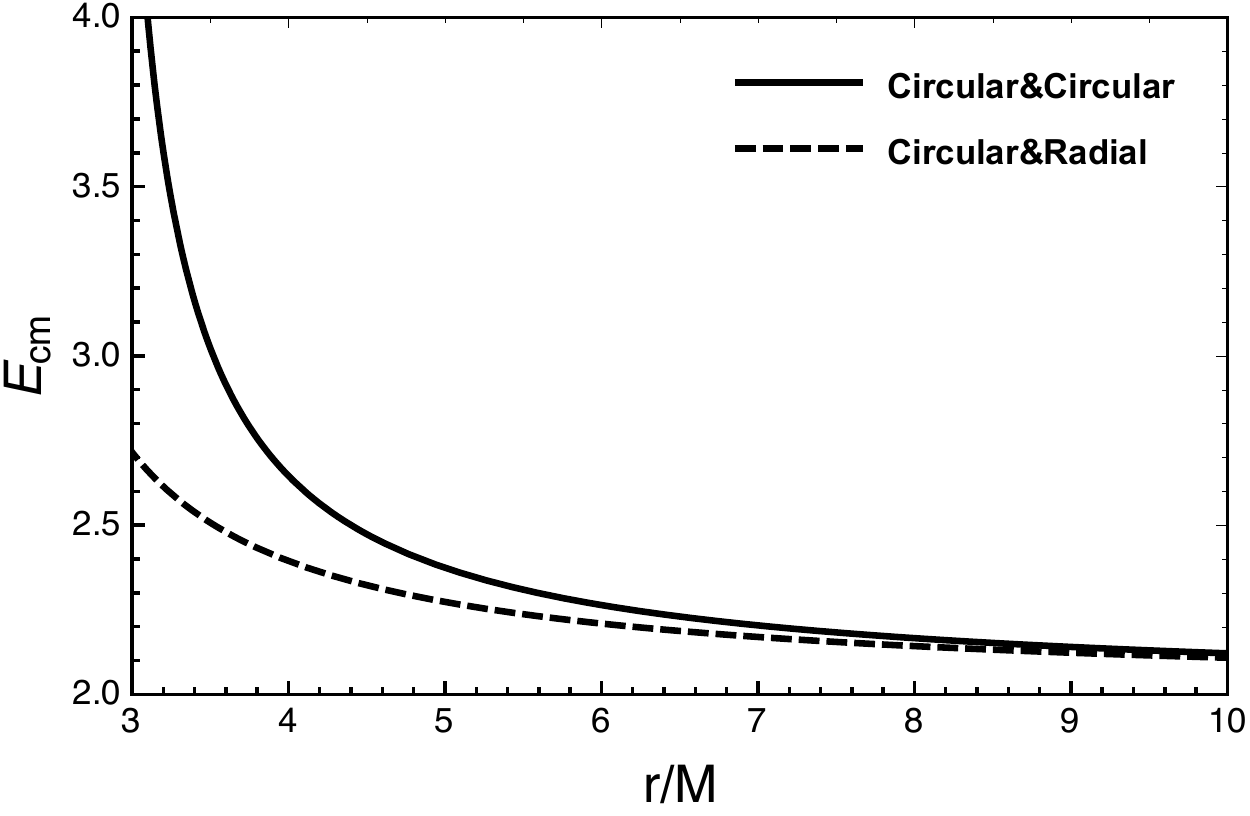}
	\includegraphics[width=0.48\columnwidth]{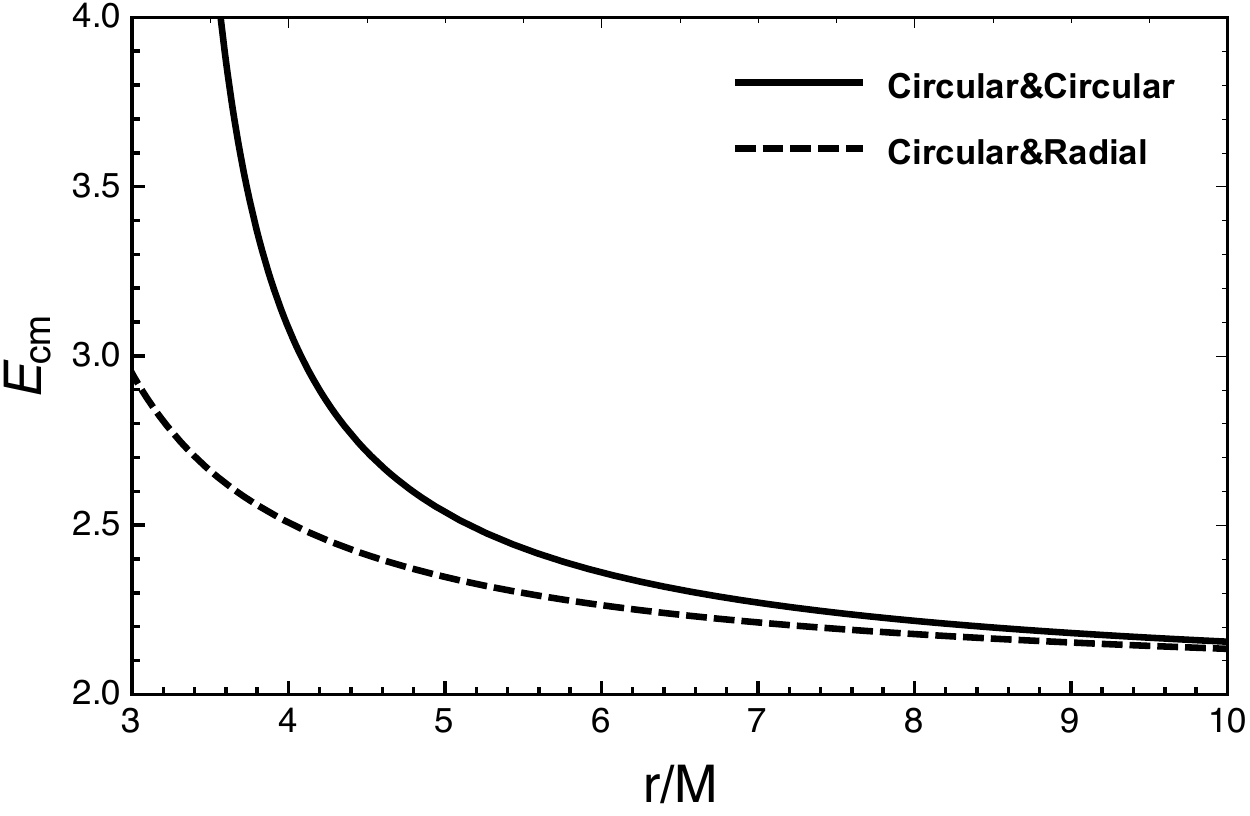}
    \caption{Collision of particles in the ZV space-time. Solid lines represent both particles on circular orbits; dashed lines represent one particle is on circular orbits and one particle falling radially from infinity. Left panel: Prolate case ($\gamma=0.9$). Right panel: Oblate case ($\gamma=1.1$).}
    \label{fig:ZVcircL1L2}
\end{figure}

It is important to notice here that not all possible orbits may occur around astrophysical compact objects. In fact collisions in the vicinity of the infinitely red-shifted surface may be extremely rare. However, there are several configurations that may be of interest, in particular, the ones involving particles on circular orbits. In general, circular orbits are not allowed in the vicinity of $r=2M$ and it is important to distinguish the region where stable circular orbits can occur from the region where circular orbits are unstable.
In the following we will consider two simple models for collisions:
\begin{enumerate}
    \item[(A)] Collision between two particles on circular orbits. In this case we consider two particles on circular orbit, i.e. $\dot{r}_1=\dot{r}_2=0$, with the same value of $r_1=r_2=r$ and opposite angular momentum $L_1=-L_2=L$.
    \item[(B)] Collision between one particle on circular orbit and one particle on radial infall from infinity. In this case we consider one particle on circular orbit, i.e. $\dot{r}_1=0$ at a radius $r_1=r$ with angular momentum $L_1=L$ and one particle on radial trajectory $r_2(\tau)$ with $L_2=0$.
\end{enumerate}
In figure \ref{fig:ERcirc11} we show the center of mass energies for the above collisions in the ER metric, while figure \ref{fig:ZVcircL1L2} shows the corresponding energies for the ZV metric. Notice that for a given radius, the collision of two particles on circular orbits with opposite angular momentum produces more energy with respect to the collision of one particle on circular orbit with a particle infalling from infinity.

\begin{figure}
	\includegraphics[width=0.48\columnwidth]{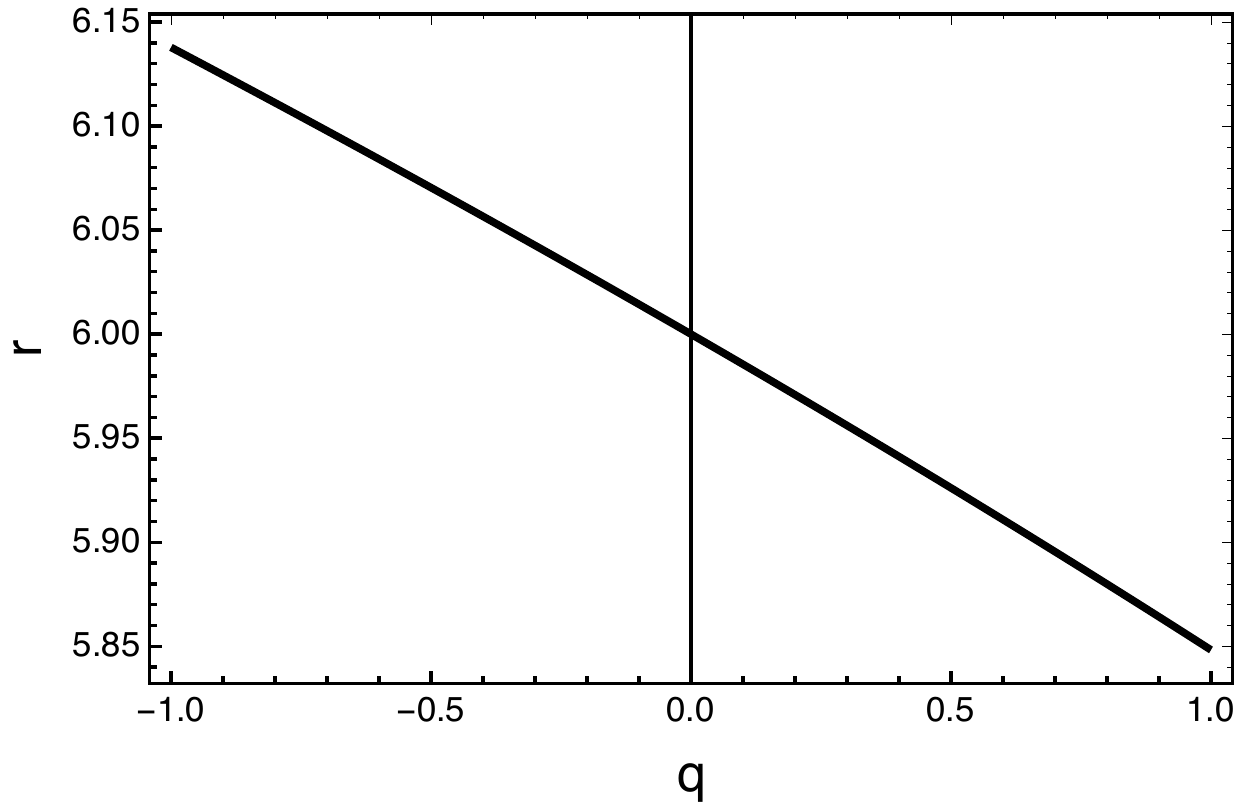}
	\includegraphics[width=0.48\columnwidth]{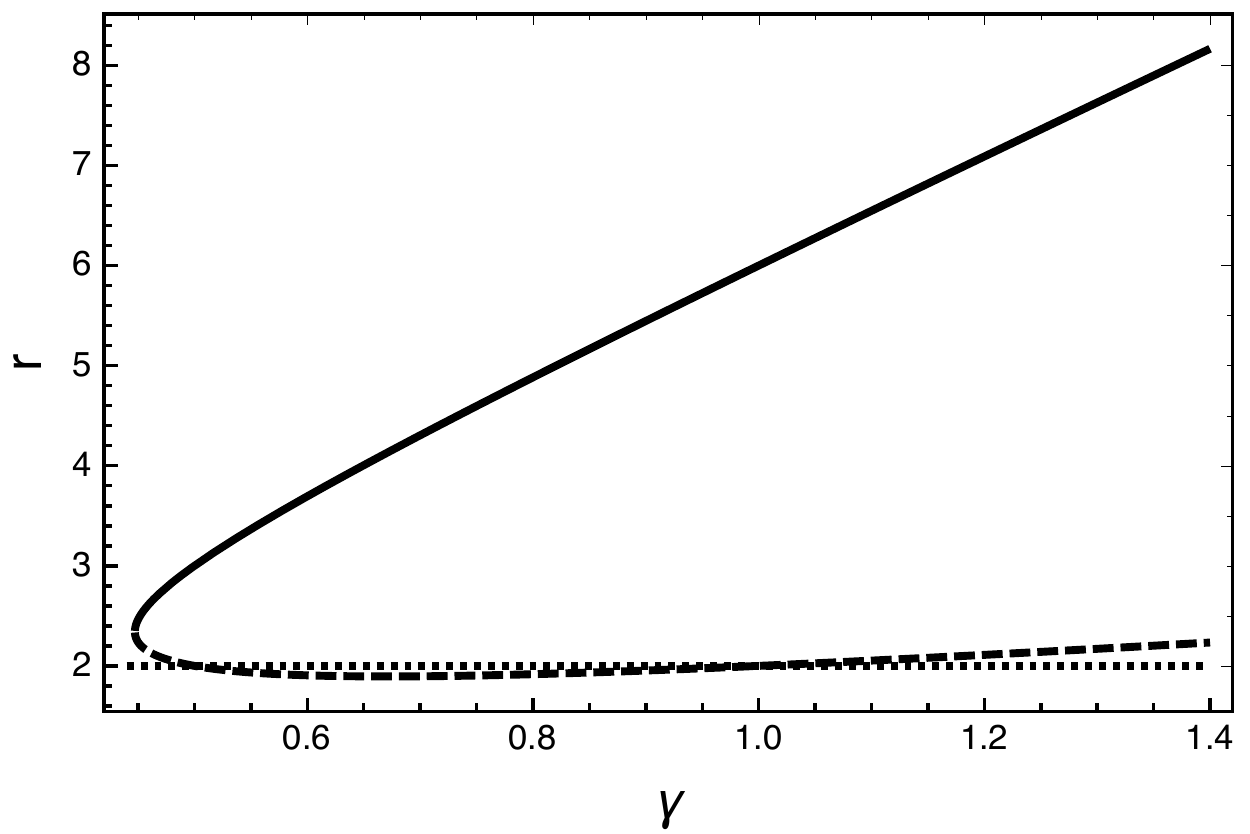}
    \caption{The dependence of the ISCO radius on the deformation parameters $q$ for ER metric (left panel) and $\gamma$ for ZV metric (right panel).}
    \label{fig:rvsq}
\end{figure}

From the point of view of astrophysics we know that compact objects are generically accompanied by accretion disks of gas falling onto them. The trajectories of particles in the accretion disk can be described as circular geodesics. However, as said, for black holes and extremely collapsed objects particles can not exist on stable circular orbits arbitrarily close to the horizon or the surface of the object. The innermost stable circular orbit (ISCO) is a common tool to probe the behaviour test particles in accretion disks near compact objects. In the case of Schwarzschild the ISCO is located at $r_{isco}=6M$, while unstable and unbound circular orbits can extend at most until $r=3M$. On the other hand for the ER and ZV metric particles can exist on stable circular orbits closer or farther from the source based on the value of the quadrupole moment. 
For a metric of Weyl's class, the effective potential for a particle on circular orbit in the equatorial plane is obtained from equation \eqref{p} setting $\dot{r}=0$ and it is given by 
\be 
U_{eff} = \frac{F^2L^2}{\Delta r^2}+F\; .
\ee
The condition for the particle to be on circular orbit is then given by $U_{eff}'=0$, where the prime denotes derivatives with respect to $r$. The energy $E$, angular momentum $L$ and angular velocity $\Omega$ per unit mass for particles on circular orbits can be also obtained from
\bea 
E^2&=&-\frac{g_{tt}^2}{g_{tt}+g_{\phi\phi}\Omega^2}\; , \\
L^2&=&-\frac{g_{\phi\phi}^2\Omega^2}{g_{tt}+g_{\phi\phi}\Omega^2}\; , \\
\Omega^2&=&-\frac{g_{tt,r}}{g_{\phi\phi,r}}\; ,
\eea
and so, for metrics of Weyl's class, we can rewrite $E$ and $L$ for particles on circular orbit as
\bea 
E^2&=&\frac{F^2L^2}{\Delta r^2}+F\; ,\\ \label{L}
L^2&=&\frac{F'\Delta^2 r^3}{F^2(\Delta' r+2\Delta)-2F'F\Delta r}\; .
\eea
The radius of the ISCO is then given by the condition of marginal stability of the orbit, i.e. $U_{eff}''=0$. 
For the ER metric the explicit form of $r_{isco}$ is rather complicated, and it is given by the solution of the following implicit function:
\be \label{ERisco}
H(r,q)=r-6+qr\Psi'(3r^2-18r+28)+qr^2\Psi''(r-2)(r-1)-6q^2r^2\Psi'^2(r-2)(r-3)+4q^3r^3\Psi'^3(r-2)^2=0 \; ,
\ee 
where for simplicity we have replaced $r\rightarrow r/M$ and where $\Psi(r)$ is given by equation \eqref{psi} with $\theta=\pi/2$.
On the other hand, for the ZV metric the ISCO radius is simply given by
\be \label{ZVisco}
r_{isco}=M(1+3\gamma\pm\sqrt{5\gamma^2-1}) \; .
\ee
Notice that there are two allowed values of $r_{isco}>2M$ for $\gamma\in(1/\sqrt{5},1/2)$. In the following we will consider only the outer radius of the ISCO thus restricting the attention to the solution with plus sign in equation \eqref{ZVisco}. In figure \ref{fig:rvsq} we show the value of $r_{isco}$ in the ER and ZV metrics as functions of $q$ and $\gamma$ respectively.
\begin{figure}
	\includegraphics[width=0.48\columnwidth]{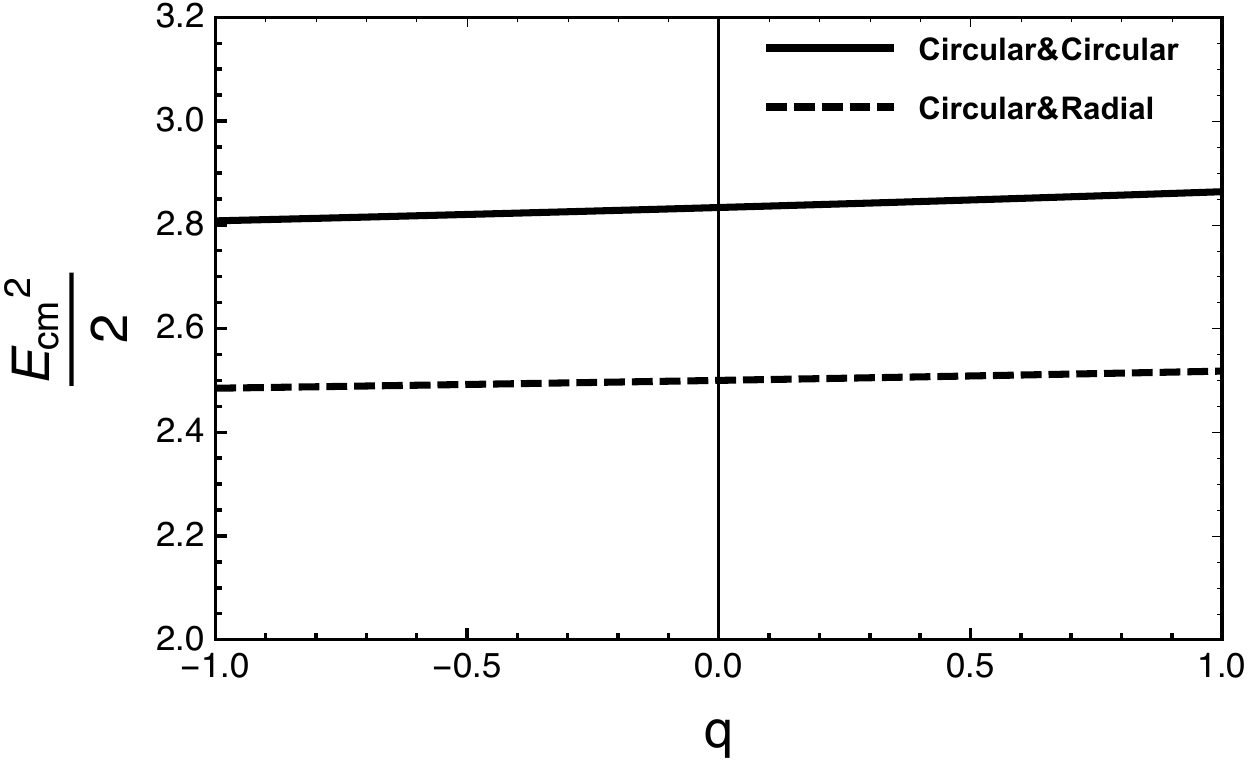}
	\includegraphics[width=0.48\columnwidth]{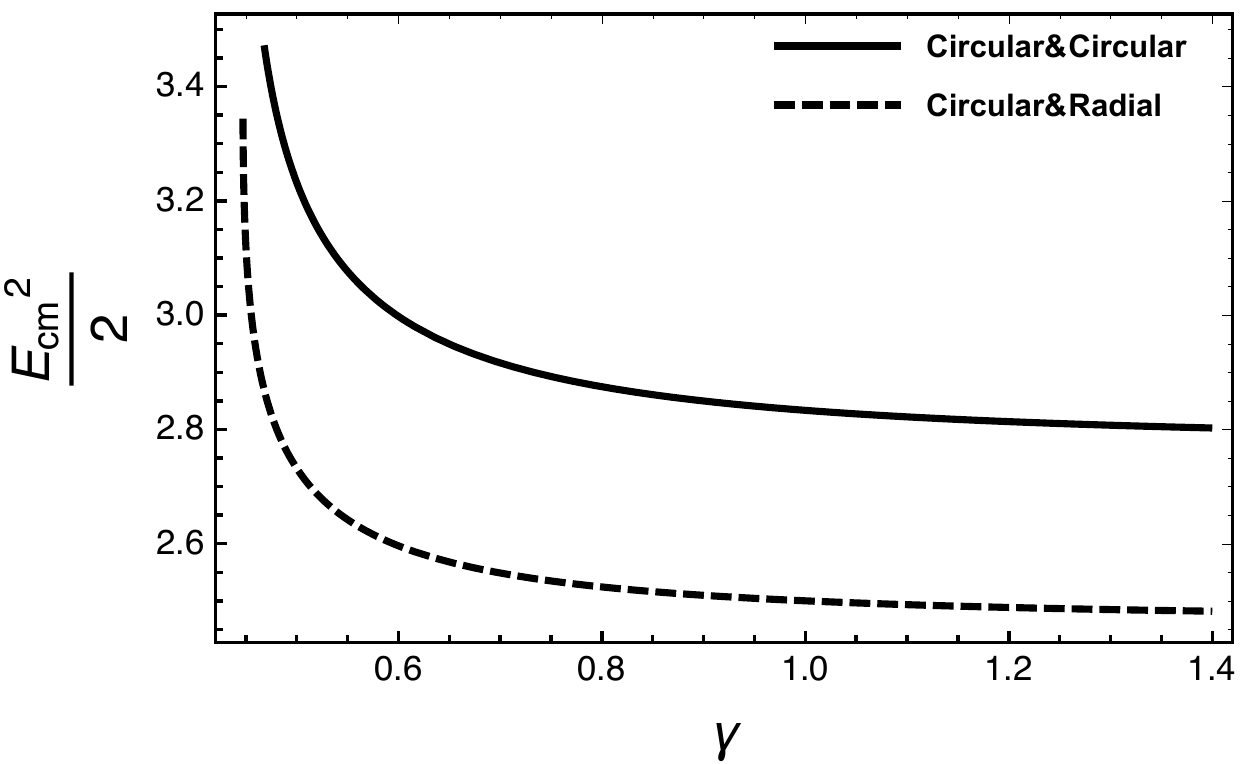}
    \caption{The center of mass energy $E_{cm}^2/2$ for collisions at the ISCO of two particles on circular orbits (solid line) and one particle on circular orbit vs. one particle on radial infall (dashed line) as a function of $q$ in the ER metric (left panel) and as a function of $\gamma$ in the ZV metric (right panel).}
    \label{fig:Evsqgamma}
\end{figure}
A word of caution must be added regarding the interpretation of the parameters $q$ and $\gamma$ in relation to physical observables. For values of $q\simeq 0$ and $\gamma\simeq 1$ we can confidently describe small departures from spherical symmetry. On the other hand, for larger values, the contribution of the quadrupole moment (or higher multipole moments) may become dominant over the mass monopole, and this may in turn affect the behavior of the orbits. For example, in the ZV metric, where monopole and the quadrupole moments contain both $M$ and $\gamma$ (see equations \eqref{ZVM0} and \eqref{ZVM2}), rewriting the equation for $r_{isco}$ in terms of $M_0$ and $M_2$ shows that there is a limiting value $M_2=4M_0^3/3$ at which the contributions of the quadrupole moment produce substantial departures from the Schwarzschild-like behavior.
A similar situation occurs in the ER metric where $M_2$ depends on both $q$ and $M$.

We now wish to evaluate the center of mass energy for particles on the accretion disk at the ISCO as a function of the deformation parameters $q$ and $\gamma$. In particular we shall restrict to the outer ISCO radius, which is the relevant case for astrophysical accretion disks.
To obtain the value of the center of mass energy for collision of two particles on circular orbits at the ISCO we use equation \eqref{Ecm} with $\dot{r}_1=\dot{r}_2=0$ and $L_1=-L_2=L$ given by equation \eqref{L}. The value of $r_{isco}$ is obtained from $U_{eff}''=0$. We then get
\be 
\frac{(E_{\rm cm}^{\rm cc})^2}{2}=1+\frac{1}{F}+\frac{F}{\Delta}\frac{L^2}{r^2} \; ,
\ee 
where the superscript `${\rm cc}$' stands for `circular-circular'.
Similarly, to obtain the center of mass energy for the collision between one particle on circular orbit at the ISCO and one on radial infall from infinity, we use equation \eqref{Ecm} with $\dot{r}_1=0$ and $L_2=0$. We obtain
\be 
\frac{(E_{\rm cm}^{\rm cr})^2}{2}=1+\frac{1}{F} \; ,
\ee 
where the superscript `${\rm cr}$' stands for `circular-radial'.

In the case of the ER metric the form of $E_{\rm cm}$ at the ISCO for the two cases is rather complicated due to the fact that the ISCO is defined via the implicit function \eqref{ERisco}. In the left panel of figure \ref{fig:Evsqgamma} we show the center of mass energy for collision of two particles at the ISCO in the ER metric in the case of both particles on circular orbits with opposite angular momentum (`cc') and one particle on circular orbit and one particle radially falling from infinity (`cr').

In the case of the ZV metric $E_{\rm cm}$ at the ISCO for the two kinds of collisions can be evaluated explicitly and it becomes
\bea
\frac{(E_{\rm cm}^{\rm cc})^2}{2}&=&1+\left(\frac{3\gamma+1+\sqrt{5\gamma^2-1}}{3\gamma-1+\sqrt{5\gamma^2-1}}\right)^\gamma+\frac{\gamma}{\gamma+\sqrt{5\gamma^2-1}} \; , \\
\frac{(E_{\rm cm}^{\rm cr})^2}{2}&=&1+\left(\frac{3\gamma+1+\sqrt{5\gamma^2-1}}{3\gamma-1+\sqrt{5\gamma^2-1}}\right)^\gamma \; .
\eea
It is easy to notice that the center of mass energy at the ISCO is greater than zero and finite for all values of $\gamma$ and it is greatest for the limiting case $\gamma=1/\sqrt{5}$ while it tends to the smallest value for $\gamma\rightarrow+\infty$.
We have the following limits
\bea 
\lim_{\gamma\rightarrow 1/\sqrt{5}} \frac{(E_{\rm cm}^{\rm cc})^2}{2}&=& 2+ \left(\frac{3+\sqrt{5}}{3-\sqrt{5}}\right)^{1/\sqrt{5}}  \; , \\
\lim_{\gamma\rightarrow 1/\sqrt{5}} \frac{(E_{\rm cm}^{\rm cr})^2}{2}&=& 1+ \left(\frac{3+\sqrt{5}}{3-\sqrt{5}}\right)^{1/\sqrt{5}} \; ,
\eea 
and
\bea 
\lim_{\gamma\rightarrow +\infty} \frac{(E_{\rm cm}^{\rm cc})^2}{2}&=& \frac{2+\sqrt{5}}{1+\sqrt{5}} \; , \\
\lim_{\gamma\rightarrow +\infty} \frac{(E_{\rm cm}^{\rm cr})^2}{2}&=& 1\; .
\eea 
It is also worth noticing that although the above limits for $\gamma\rightarrow 1/\sqrt{5}$ are finite the function $f(\gamma)$ given by $E_{\rm cm}^2/2$ evaluated at $r_{isco}$ approaches the limit with infinite tangent in both cases, i.e. $df/d\gamma\rightarrow+\infty$ for $\gamma\rightarrow 1/\sqrt{5}$. The center of mass energies for the two scenarios in the ZV metric are shown in the right panel of figure \ref{fig:Evsqgamma}. The difference between the same scenarios in the ER metric with respect to the ZV metric is due to the dependence of the $r_{isco}$ on the deformation parameter for the two metrics. In fact the ISCO in the ZV space-time approaches $r=2M$ as $\gamma\rightarrow 1/\sqrt{5}$ causing the faster blow up of the center of mass energy, while for the ER metric the ISCO does not depart significantly from the Schwarzschild value $r=6M$ even for values of $q$ close to zero.

Finally, it is worth mentioning that the above analysis is constrained to the equatorial plane $\theta=\pi/2$. However, strictly speaking, by considering motion restricted to the equatorial plane one may not be able to distinguish a vacuum space-time with quadrupole moment from a non vacuum spherically symmetric space-time. More precisely a metric for which $-g_{tt}=1/g_{RR}=F(R(r),\pi/2)$ with $R(r)$ chosen in such a way that $g_{\phi\phi}=R^2\sin^2\theta$ would exhibit the same properties for motion of particles as the line element in equation \eqref{line} when $\theta=\pi/2$. Nevertheless, the matter distribution producing such line element would not necessarily be physically valid and the interpretation of a vacuum space-time with quadrupole moment appears to be more natural. 

\begin{figure}
	\includegraphics[width=0.48\columnwidth]{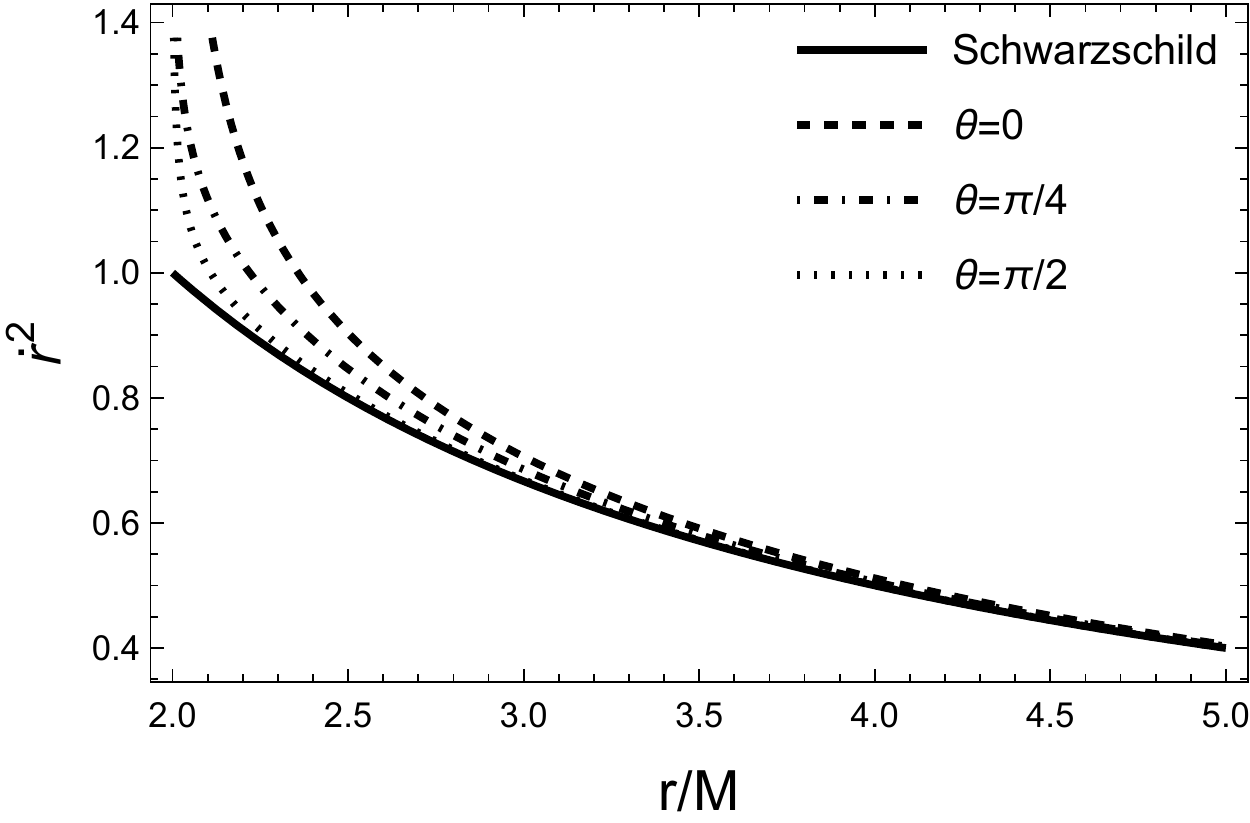}
	\includegraphics[width=0.48\columnwidth]{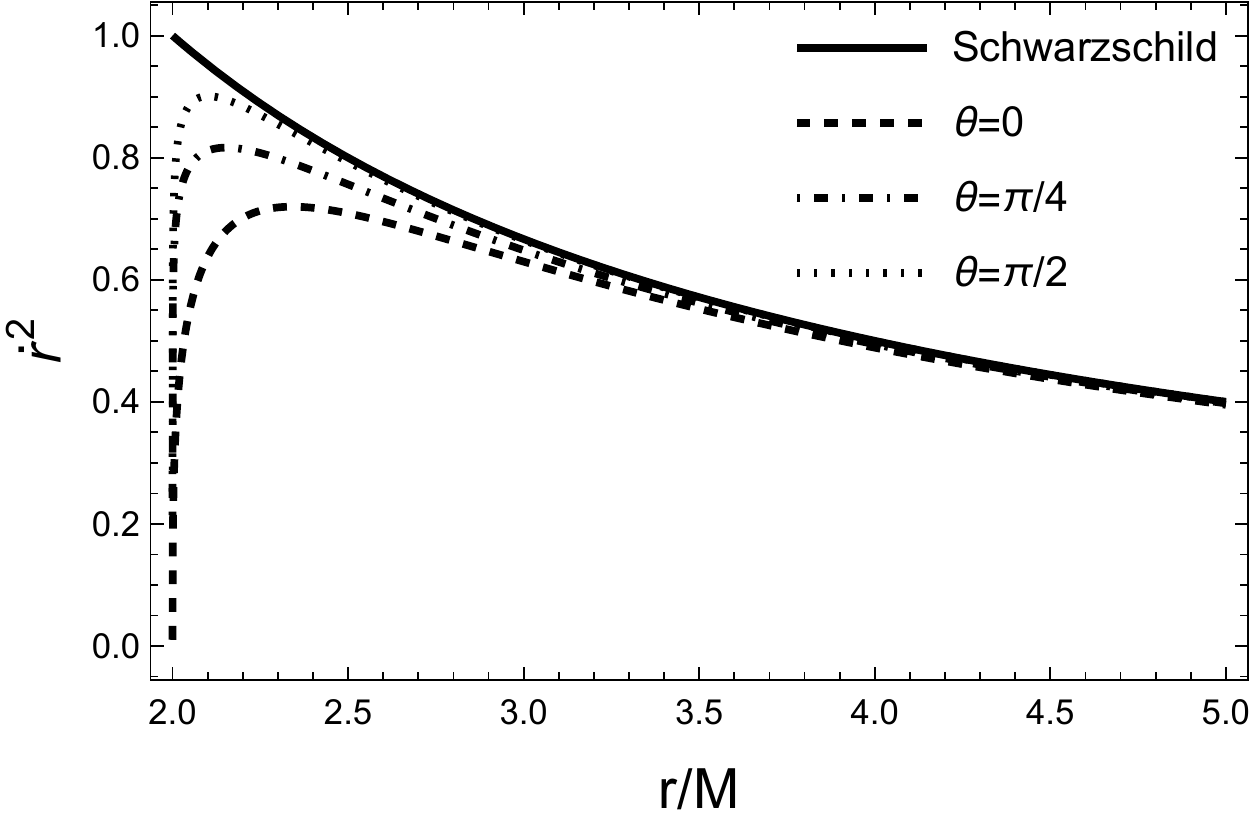}
    \caption{The square of the radial velocity for a test particle in radial fall from infinity in the Erez-Rosen space-time for different values of the latitudinal angle $\theta$ for oblate (left panel, with $q=-0.1$) and prolate (right panel, with $q=0.1$) sources.}
    \label{fig:theta-ER}
\end{figure}

\begin{figure}
	\includegraphics[width=0.48\columnwidth]{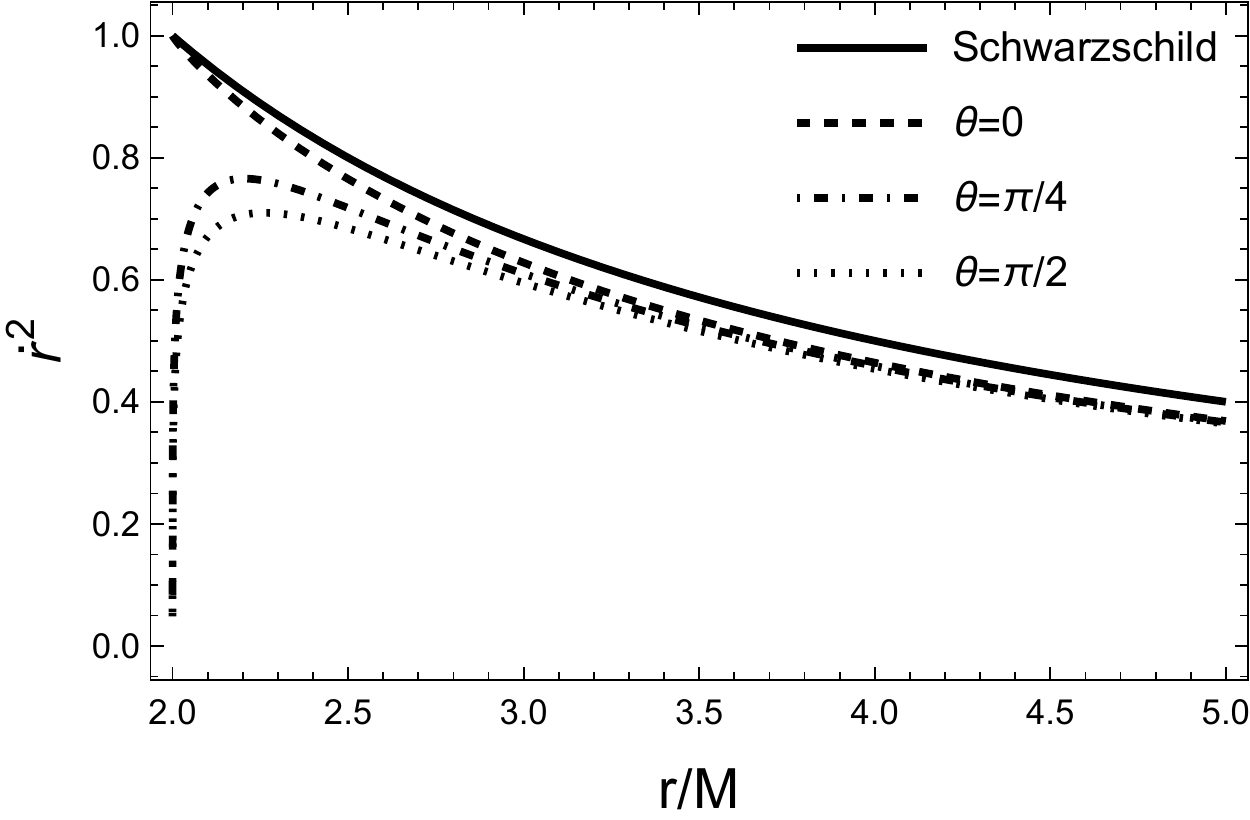}
	\includegraphics[width=0.48\columnwidth]{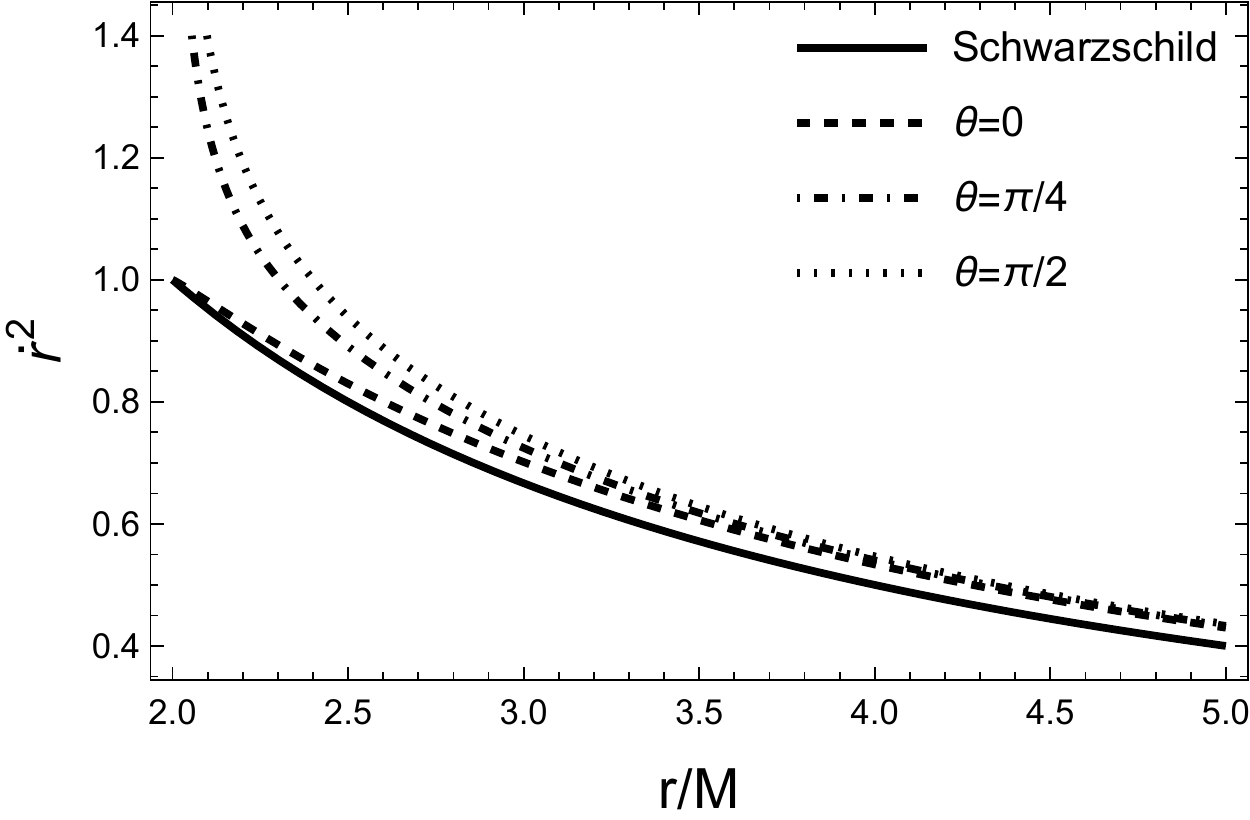}
    \caption{The square of the radial velocity for a test particle in radial fall from infinity in the Zipoy-Voorhees space-time for different values of the latitudinal angle $\theta$ for prolate (left panel, with $\gamma=1.1$) and oblate (right panel, with $\gamma=0.9$) sources.}
    \label{fig:theta-ZV}
\end{figure}

Of course, the study of the motion of test particles outside the equatorial plane is more complicated with respect to the equatorial case and when dealing with accretion disks around compact objects one can expect particles to orbit close to the equatorial plane. For completeness, we considered the effects of the latitudinal angle $\theta$ on the velocity of test particles infalling radially for infinity. These are shown in figure \ref{fig:theta-ER} for the ER metric and figure \ref{fig:theta-ZV} for the ZV metric.
As expected, the velocities are larger with respect to Schwarzschild for oblate sources and smaller for prolate sources and the same behavior is expected to translate in the collision energy. Nevertheless it is worth noticing that while the behavior for the ER metric becomes closer to Schwarzschild for particles closer to the equatorial plane, the opposite happens in the ZV metric. This is due to the term $M^2\sin^2\theta/r^2$ in the metric function $g_{rr}$ in equation \eqref{ZV-grr} which vanishes for $\theta=0$ thus giving exactly $-g_{tt}=1/g_{rr}=\Delta^\gamma$ along the symmetry axis.


\section{Discussion}\label{discussion}

Determining from observations whether an astrophysical compact object is indeed a black hole is one of the main challenges to modern relativistic astrophysics. Black hole candidates in the universe may be in binary systems or immersed in external matter and electromagnetic fields (see for example \cite{dist, LL, BMN} for solutions describing black holes surrounded by other fields). In addition, such objects may differ from isolated black holes as they may have a more complicated multipolar structure, such as the ones considered in this paper. Therefore determining to what extent observations are able to distinguish different geometries is of crucial importance.

We considered the collision of test particles in the gravitational field of black hole mimickers described by exact solution of the vacuum Einstein's field equations with non vanishing quadrupole moment. As expected, with the exception of the special case of spherical symmetry (i.e. Schwarzschild), the center of mass energy for collisions is defined only outside the infinitely red-shifted surface $r=2M$ which acts as the lower allowed limit for the object's boundary. Under the assumption that exotic compact sources with quadrupole moment may exist in nature, our results show that particle collisions in close proximity of the surface of prolate objects may produce arbitrarily high energies. However, the collision energy at the innermost stable circular orbit is finite for both oblate and prolate objects and depends strongly on the value of the quadrupole moment.
Therefore, at least in principle, such collisions could be used as a way to distinguish these kinds of exotic sources from black holes and thus test the possibility of existence of extreme compact objects with quadrupole moment in the universe.

\section*{Acknowledgement}
The work was supported by Nazarbayev University Faculty Development Competitive Research Grant No. 090118FD5348 and by the Ministry of Education and Science of the Republic of Kazakhstan target program IRN: BR05236454.


\begin{thebibliography}{99}

\bibitem{NH1}	W.~Israel,
	%``Event horizons in static vacuum space-times,''
	Phys. Rev. \textbf{164}, 1776 (1967).
%	doi:10.1103/PhysRev.164.1776

\bibitem{NH2}	B.~Carter,
	%``Axisymmetric Black Hole Has Only Two Degrees of Freedom,''
	Phys. Rev. Lett. \textbf{26}, 331 (1971).
%	doi:10.1103/PhysRevLett.26.331

\bibitem{gur}	N.~G\"urlebeck,
	%``No-hair theorem for Black Holes in Astrophysical Environments,''
	Phys. Rev. Lett. \textbf{114} (15), 151102 (2015).
%	doi:10.1103/PhysRevLett.114.151102
%	[arXiv:1503.03240 [gr-qc]].

\bibitem{herrera} L.~Herrera,
%The Israel Theorem: What is Nature Trying to Tell Us?
Int. J. Mod. Phys. D {\bf 17}, 557 (2008).

\bibitem{Bonnor1992}	W.~B.~Bonnor,
	%``Physical interpretation of vacuum solutions of Einstein's equations. Part I. Time-independent solutions,''
	Gen. Relat. Gravit. {\bf24}, 551 (1992).
 %	https://doi.org/10.1007/BF00760137

\bibitem{weyl} H.~Weyl, 
Ann. Phys. {\bf 54}, 117 (1917). 

\bibitem{weyl2} H.~Weyl, 
Ann. Phys. {\bf 59}, 185 (1919).

\bibitem{Quevedo1990}	H.~Quevedo,
	%``Multipole Moments in General Relativity—Static and Stationary Vacuum Solutions—,''
	Fortschr. Phys. {\bf 58}, 10, 733 (1990).
	
\bibitem{Pastora1994} J.~L.~Hernandez-Pastora and J.~Martin,
	Gen. Relat. Gravit. {\bf26}, 877 (1994).
	
\bibitem{sing1}
	R.~Penrose,
	%``Gravitational collapse and space-time singularities,''
	Phys. Rev. Lett. \textbf{14}, 57 (1965).
	%doi:10.1103/PhysRevLett.14.57
	%990 citations counted in INSPIRE as of 13 May 2020

\bibitem{sing2}
	S.~Hawking and R.~Penrose,
	%``The Singularities of gravitational collapse and cosmology,''
	Proc. Roy. Soc. Lond. A \textbf{A314}, 529 (1970).
	%doi:10.1098/rspa.1970.0021
	%814 citations counted in INSPIRE as of 13 May 2020
	
\bibitem{inf} P. Chen, Y. C. Ong and D.-h. Yeom,
%Black hole remnants and the information loss paradox
Phys. Rep. {\bf 603}, 1 (2015).
	
\bibitem{malafarina} 		D.~Malafarina,
	%``Classical collapse to black holes and quantum bounces: A review,''
	Universe \textbf{3} (2), 48 (2017).
%	doi:10.3390/universe3020048
%	[arXiv:1703.04138 [gr-qc]].

\bibitem{orb1} 
		A.~N.~Chowdhury, M.~Patil, D.~Malafarina and P.~S.~Joshi,
		%``Circular geodesics and accretion disks in Janis-Newman-Winicour and Gamma metric,''
		Phys.\ Rev.\ D {\bf 85}, 104031 (2012).
%		doi:10.1103/PhysRevD.85.104031
%		[arXiv:1112.2522 [gr-qc]].

\bibitem{orb2}
	K.~Boshkayev, E.~Gasperin, A.~Gutierrez-Pineres, H.~Quevedo and S.~Toktarbay,
	%``Motion of test particles in the field of a naked singularity,''
	Phys. Rev. D \textbf{93} (2), 024024 (2016).
%	doi:10.1103/PhysRevD.93.024024
%	[arXiv:1509.03827 [gr-qc]].

\bibitem{ER1} G. Erez and N. Rosen,
Bull. Res. Council Israel {\bf 8F}, 47 (1959).

\bibitem{armenti1977}
A. Armenti, 
Int. J. Theor. Phys. {\bf 16} (11), 813 (1977).

\bibitem{ER2} 
D. Bini, M. Crosta, F. de Felice, A. Geralico and A. Vecchiato,
%The Erez-Rosen metric and the role of the quadrupole on light propagation
Class. Quantum Grav. {\bf 30}, 045009 (2013).

\bibitem{ER3} H. Quevedo and L. Parker,
	Gen. Relat. Gravit. {\bf 23}, 495 (1991).

\bibitem{ER4} K. Boshkayev, H. Quevedo, G. Nurbakyt, A. Malybayev and A. Urazalina,
%The Erez–Rosen Solution Versus the Hartle–Thorne Solution
Symmetry, {\bf 11} (10), 1324 (2019).


\bibitem{ZV1}
	D.~M.~Zipoy
%	``Static axially symmetric gravitational fields,''
	Journ. Math. Phys. {\bf 22}, 1137 (1970).
%	https://doi.org/10.1063/1.1705005

\bibitem{ZV2}
	B.~H.~Voorhees,
%	``Static axially symmetric gravitational fields,''
	Phys.\ Rev.\ D {\bf 2}, 2119 (1970).
%	doi:10.1103/PhysRevD.2.2119
	
\bibitem{ZV3}
	D.~Papadopoulos, B.~Stewart and L.~Witten,
	%``Some Properties of a Particular Static, Axially Symmetric Space-time,''
	Phys. Rev. D \textbf{24}, 320 (1981).
%	doi:10.1103/PhysRevD.24.320
	
	\bibitem{ZV4}
	L.~Herrera, F.~M.~Paiva and N.~Santos,
	%``Geodesics in the gamma space-time,''
	Int. J. Mod. Phys. D \textbf{9}, 649 (2000).
%	doi:10.1142/S021827180000061X
	
\bibitem{orb3} 
		C.~A.~Benavides-Gallego, A.~Abdujabbarov, D.~Malafarina, B.~Ahmedov and C.~Bambi,
		%``Charged particle motion and electromagnetic field in $\gamma$ space-time,''
		Phys.\ Rev.\ D {\bf 99} (4), 044012 (2019).
%		doi:10.1103/PhysRevD.99.044012
%		[arXiv:1812.04846 [gr-qc]].
				
\bibitem{orb4}
	A.~B.~Abdikamalov, A.~A.~Abdujabbarov, D.~Ayzenberg, D.~Malafarina, C.~Bambi and B.~Ahmedov,
	%``Black hole mimicker hiding in the shadow: Optical properties of the $\gamma$ metric,''
	Phys. Rev. D \textbf{100} (2), 024014 (2019).
%	doi:10.1103/PhysRevD.100.024014
%	[arXiv:1904.06207 [gr-qc]].	

\bibitem{orb5}
	B.~Toshmatov, D.~Malafarina and N.~Dadhich,
	%``Harmonic oscillations of neutral particles in the $\gamma$-metric,''
	Phys. Rev. D \textbf{100} (4), 044001 (2019).
%	doi:10.1103/PhysRevD.100.044001
%	[arXiv:1905.01088 [gr-qc]].

\bibitem{orb6}
		B.~Toshmatov and D.~Malafarina,
		%``Spinning test particles in the $\gamma$ spacetime,''
		Phys. Rev. D \textbf{100} (10), 104052 (2019).
%		doi:10.1103/PhysRevD.100.104052
%		[arXiv:1910.11565 [gr-qc]].

\bibitem{BSW} M. Ba\~nados, J. Silk and S. M. West, 
%Kerr Black Holes as Particle Accelerators to Arbitrarily High Energy, 
Phys. Rev. Lett. {\bf 103} (11), 111102 (2009).

\bibitem{zav} O. B. Zaslavskii,
%High energy particle collisions and geometry of horizon
Int. J. Mod. Phys. D
{\bf 25} (11), 1650095 (2016).

\bibitem{coll1} T. Jacobson and T. P. Sotiriou, 
%Spinning Black Holes as Particle Accelerators,
Phys. Rev. Lett. {\bf 104} (2), 021101 (2010).

\bibitem{coll2} T. Harada and M. Kimura, 
%Collision of an innermost stable circular orbit particle around a Kerr black hole, 
Phys. Rev. D {\bf 83} (2), 024002 (2011).

\bibitem{coll3} O. B. Zaslavskii, 
%Acceleration of particles as a universal property of rotating black holes, 
Phys. Rev. D {\bf 82} (8), 083004 (2010).

\bibitem{coll4} M. Kimura, K.-I. Nakao and H. Tagoshi, 
%Acceleration of colliding shells around a black hole: Validity of the test particle approximation in the Banados-Silk-West process, 
Phys. Rev. D {\bf 83} (4), 044013 (2011).

\bibitem{coll5} M. Patil and P. S. Joshi, 
%Naked singularities as particle accelerators, 
Phys. Rev. D {\bf 82} (10), 104049 (2010).

\bibitem{coll6} M. Patil, P. S. Joshi and D. Malafarina, 
%Naked singularities as particle accelerators. II., 
Phys. Rev. D {\bf 83} (6), 064007 (2011).

\bibitem{young} J. H. Yound and C. A. Coulter,
%Exact Metric for a Nonrotating Mass with a Quadrupole Moment
Phys. Rev, {\bf 184}, 1313 (1969).

\bibitem{curzon} H. E. J. Curzon, 
Proc. London Math. Soc. {\bf 23}, 477 (1924).

%\bibitem{semerak} 
%O. Semer\'ak, T. Zellerin and M. Z\'acek,
%Mon. Not. R. Astron. Soc. {\bf 308}, 691 (1999).
%The structure of superposed Weyl fields

\bibitem{multipoles1} R. Geroch, J. Math. Phys. {\bf 11}, 2580 (1970). 
\bibitem{multipoles2} R. O. Hansen, J. Math. Phys. {\bf 15}, 46 (1974).
\bibitem{multipoles3} K. S. Thome, Rev. Mod. Phys. {\bf 52}, 299 (1980). 
\bibitem{multipoles4} G. Fodor, C. Hoenselaers, and Z. Perj\'es, J. Math. Phys. {\bf 30}, 2252 (1989).

\bibitem{semerak}
O. Semerak, 
Phys. Rev. D {\bf 94}, 104021 (2016).

\bibitem{bini2}
D. Bini, A. Geralico and A. Passamonti,
%Radiation drag in the field of a non-spherical source
MNRAS {\bf 446}, 65 (2015). 
%doi:10.1093/mnras/stu2082

\bibitem{kodama}
H. Kodama and W. Hikida, 
Class. Quantum Grav. {\bf 20}, 5121 (2003).

\bibitem{Teukolvsky} S. A. Teukolsky, 
APJ {\bf 185}, 635 (1973).

\bibitem{HP} L. Hernandez-Pastora and J. Martin,
Class. Quantum Grav. {\bf 10}, 2581 (1993).

\bibitem{gutsunayev} Ts. I. Gutsunayev and V. S. Manko,
Gen. Relat. Gravit. {\bf 17}, 1025 (1985).

\bibitem{armenti1972} A. Armenti, 
%A classification of particle motions in the equatorial plane of a gravitational monopole-quadrupole field in Newtonian mechanics and general relativity. 
Celestial Mechanics {\bf 6}, 383 (1972).

\bibitem{dist} 
R. Geroch and J. B. Hartle, 
J. Math. Phys. {\bf 23}, 680 (1982).

\bibitem{LL} 
J. P. S. Lemos and P. S. Letelier,
%Exact general relativistic thin disks around black holes
Phys. Rev. D {\bf 49}, 5135 (1994).

\bibitem{BMN}
C. Bambi, D. Malafarina and N. Tsukamoto,
%Note on the effect of a massive accretion disk in the measurements of black hole spins
Phys. Rev. D {\bf 89}, 127302 (2014).


\end{thebibliography}
\end{document}